\documentclass[sigconf]{acmart}

\renewcommand\footnotetextcopyrightpermission[1]{}
\acmConference[ASPDAC '27]
{32nd Asia and South Pacific Design Automation Conference}
{January 25--28, 2027}
{Tokyo, Japan}

\acmYear{2027}
\copyrightyear{2027}

\usepackage{amsmath}
\usepackage{algorithmic}
\usepackage{graphicx}
\usepackage{textcomp}
\usepackage{xcolor}
\usepackage{booktabs}
\usepackage{multirow}
\usepackage{enumitem}

\newcommand{\code}[1]{\textnormal{\texttt{#1}}}


\begin{document}

\title{SpecLens: LLM-Based Verilog Generation with Specification-Derived Constraints via Behavioral Divergence}

\author{Wen Bing}
\affiliation{%
  \institution{TU Ilmenau}
  \city{Ilmenau}
  \country{Germany}
}
\email{wen.bing@tu-ilmenau.de}

\author{Bing Li}
\affiliation{%
  \institution{TU Ilmenau}
  \city{Ilmenau}
  \country{Germany}
}
\email{bing.li@tu-ilmenau.de}

\begin{abstract}
Large language models (LLMs) have recently shown promise in Verilog generation, but producing functionally correct RTL directly from natural-language specifications remains a highly challenging task. Existing approaches improve LLM-based Verilog generation mainly with retrieval-augmented generation (RAG), self-planning, or few-shot prompting. However, these methods focus primarily on external or generic forms of enhancement rather than strengthening the specification with task-specific constraints. In this work, we propose SpecLens, an automated framework for LLM-based Verilog generation that derives specification-driven constraints by analyzing behavioral divergence among multiple candidate implementations, using the original specification as the only external semantic source during generation. On the VerilogEval v2.0 spec-to-RTL benchmark, SpecLens achieves a functional pass@1 ratio of 86.2\% with o3-mini-medium and 89.4\% with o3-mini-high. This corresponds to a 3.6 percentage-point gain over the SOTA prompting method with o3-mini-medium and a 3.8 percentage-point gain over the SOTA behavioral divergence method with o3-mini-high. In addition, on RTLLM v1.1 and v2.0, analysis shows that SpecLens is more specification-faithful and less prone to benchmark-aligned priors. SpecLens achieves 100\% syntactic correctness on VerilogEval v2.0, 86.2\% on RTLLM v1.1, and 88\% on RTLLM v2.0, even without using costly compile-repair loops to revise generated code iteratively. The code is open source and available at \url{https://anonymous.4open.science/r/SpecLens-4632/readme.md}.

\end{abstract}

\maketitle

\section{Introduction}

Large language models (LLMs) have stimulated growing interest in their application to electronic design automation (EDA), particularly for generating register-transfer-level (RTL) designs in Verilog from natural-language specifications. Early work explores Verilog generation based on GPT~\cite{first_llm_to_verilog}, while subsequent studies investigate reinforcement learning with hardware-aware rewards~\cite{Reinforcement2}, HLS-C repair~\cite{HLS-c}, and automated testbench generation~\cite{Autobench,correctbench}. Despite this progress, functional correctness remains limited due to the scarcity of high-quality Verilog training data and model hallucinations~\cite{hallucination}.

A straightforward solution to improve correctness is to rely on testbench feedback or fine-tuning. Multi-agent approaches~\cite{MAGE,VerilogCoder} have reported strong results, but testbenches are often unavailable in practice when RTL generation starts from natural-language specifications. Fine-tuning methods~\cite{finetune2,finetune3,finetune4} have also shown strong performance on specific datasets, but there is insufficient high quality Verilog data to fine-tune a general Verilog model for broad use. This motivates an alternative direction: improving Verilog generation by enhancing the specification.

Existing specification-enhancement approaches also have important limitations. Interactive workflows~\cite{chipchat, Clarifygpt} can obtain useful information, but require substantial human effort and hardware expertise. RAG methods~\cite{rag1,rag2,rag3} attempt to strengthen specification understanding or repair Verilog syntax using external libraries. However, their effectiveness depends heavily on retrieval quality and may degrade when the retrieved examples are not sufficiently important for the current task or are misleading. In-context learning methods such as few-shot prompting provide examples that help the model become more familiar with tasks in this domain~\cite{promptengineering}. Self-planning asks the LLM to derive an implementation plan from the specification, thereby introducing additional intermediate information~\cite{RTLLM_promptengineering}. However, these methods do not dynamically provide the task-specific information that the model actually needs for each problem, especially for parts of the specification that are unfamiliar or difficult for the LLM. Therefore, the key challenge is to determine when and where a specification is ambiguous from the model's perspective.

A possible approach to identify such ambiguity is through self-consistency-based methods, in which behavioral divergence signals potential specification ambiguity. The intuition is that if a specification is sufficiently clear to the LLM, independently generated implementations should have the same behavior even across multiple sampling runs. In the software domain, prior work~\cite{Clarifygpt} shows that output inconsistency among multiple LLM-generated programs can expose specification ambiguity, but resolving such ambiguity still requires further human interaction. SpecFix~\cite{Specfix} uses golden examples provided with the specification to identify where behavioral disagreement arises. However, hardware benchmarks typically do not provide such golden examples. In the RTL domain, methods based on self-consistency such as VRank~\cite{vrank} and VFocus~\cite{vfocus} have also been explored, but they mainly use behavioral clustering as a back-end mechanism for candidate ranking or selection. As a result, they do not provide targeted semantic supplementation for those parts of the specification that the model fails to understand well.

We address this gap by using behavioral divergence not merely as a ranking signal but as a source of specification-derived disambiguation constraints. If a specification is sufficiently explicit, independently generated Verilog candidates should converge to the same observable behavior under the same stimuli. Based on this idea, we propose SpecLens, an automated constraint derivation framework for LLM-based Verilog generation. SpecLens extracts explicit behavioral requirements from the given specification, constructs structured scenarios and stimuli, analyzes Verilog output behavior, and derives disambiguation constraints. To improve reliability, it further applies entropy-guided filtering, retaining only those constraints that reduce behavioral uncertainty. 
Our key contributions are as follows:

\begin{itemize}[leftmargin=1.1em,labelsep=0.4em,itemsep=2pt,topsep=2pt]
       \item We propose SpecLens, an automated framework for LLM-based Verilog generation that derives disambiguation constraints directly from the original specification by using behavioral divergence among candidate implementations to identify ambiguities from the model's perspective, without relying on external retrieval or golden testbenches during generation.
    \item On VerilogEval v2.0 spec-to-RTL, SpecLens achieves 100\% syntactic pass@1 with o3-mini-medium without using any syntax repair loop during generation, and also reaches the best functional pass@1 among all compared baselines, achieving 86.2\% with o3-mini-medium and 89.4\% with o3-mini-high.
    \item The derived constraints are reusable across models and can serve as additional guidance for smaller models. With these constraints, GPT-4o-mini improves functional pass@1 by 6.84\% compared to using the original specification alone.
    \item Experiments on RTLLM show that SpecLens is more sensitive to explicit local changes to the specification and therefore produces more specification-faithful implementations, while other direct generation methods tend to reproduce previously learned canonical implementations, which can introduce critical discrepancies in the results.
\end{itemize}

\begin{figure}[t]
    \centering
    \includegraphics[width=\columnwidth]{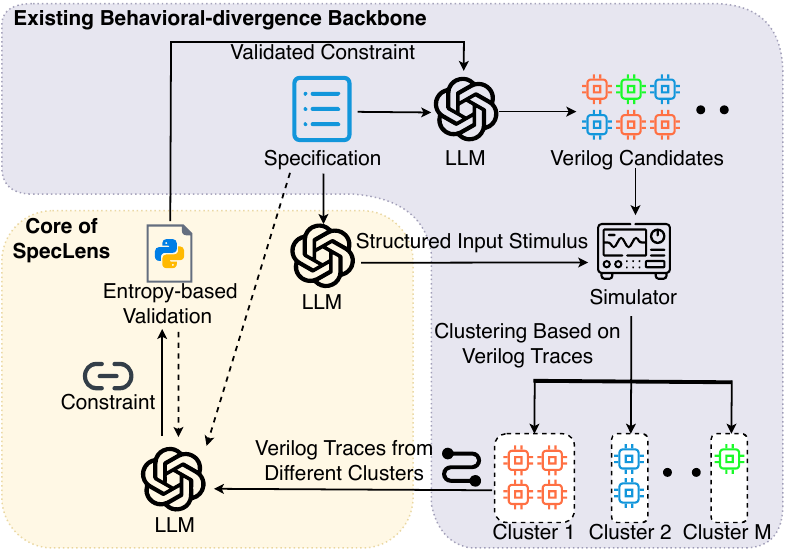}
    \Description{Core of SpecLens.}
    \caption{Core of SpecLens.}
    \label{fig:Core_of_SpecLens}
\end{figure}

\vspace{-4pt}
\section{Concept Overview}
 In this section, we briefly introduce the fundamental changes made by SpecLens over existing behavioral-divergence-based workflows and clarify several concepts used throughout the rest of the paper. As illustrated in Figure~\ref{fig:Core_of_SpecLens}, existing behavioral-divergence backbones mainly apply input stimuli to multiple LLM-generated Verilog candidates and then select the final Verilog from the resulting behavioral clusters. However, they do not systematically construct or exploit the information provided by behavioral clustering.

The core of SpecLens lies in three aspects: how LLMs generate structured input stimuli, how LLMs derive constraints from behavioral clustering, and how the framework validates whether the derived constraints effectively drive subsequent Verilog generation toward convergence. In all figures in this paper, solid arrows indicate the main execution flow, whereas dashed arrows denote contextual data dependencies. In this work, a Verilog trace refers to the sampled Verilog output sequence produced by a Verilog implementation in the simulator under a given input stimulus. Behavioral divergence refers to the case where different Verilog candidates produce different Verilog traces under the same stimulus. Candidates with identical Verilog traces are grouped into the same cluster. The process of applying a stimulus to Verilog candidates and partitioning them into different clusters is referred to as clustering.

\begin{figure}[t]
    \centering
    \includegraphics[width=\columnwidth]{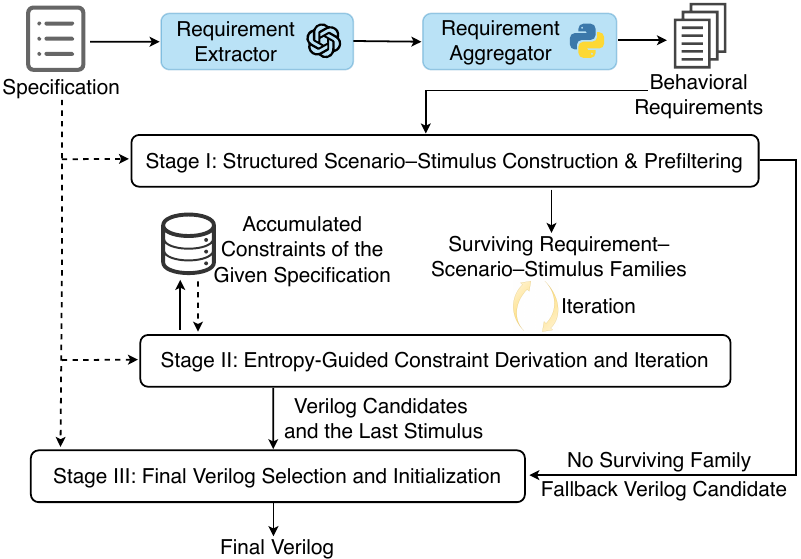}
    \Description{Overall workflow of SpecLens.}
    \caption{Overall workflow of SpecLens.}
    \label{fig:framework_overview}
\end{figure}

\begin{figure*}[t]
    \centering
    \includegraphics[width=\textwidth]{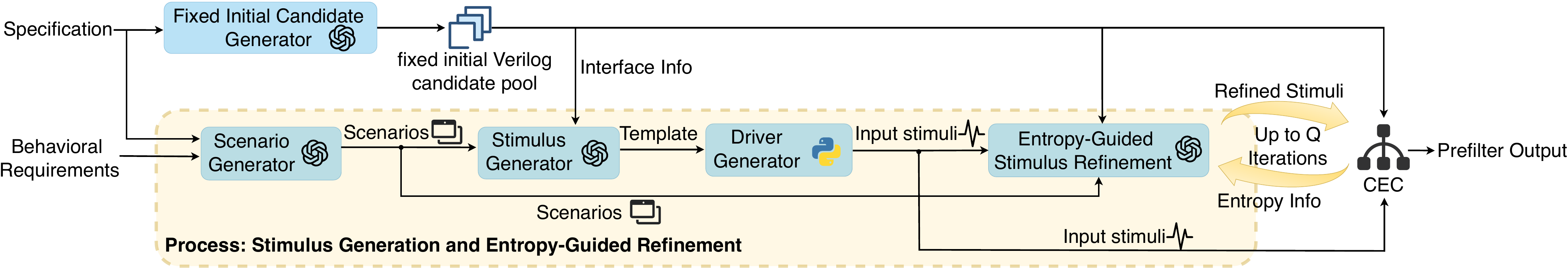}
    \Description{Stage~I: Structured scenario--stimulus construction and prefiltering.}
    \caption{Stage~I: structured scenario--stimulus construction and prefiltering.}
    \label{fig:stage1_prefilter}
\end{figure*}

\section{Methodology}
\subsection{Framework Overview}

In this section, we present the overall workflow of SpecLens, as shown in Figure~\ref{fig:framework_overview}. Starting from the original specification, the framework first extracts behavioral requirements describing expected design behavior rather than interface definitions. A Python-based Requirement Aggregator concatenates all extracted behavioral requirements into an additional composite requirement, which is appended to the requirement list and processed after the individual requirements. The stimulus generated from this composite requirement serves as a global behavioral evaluation stimulus for final candidate selection in Stage~III, since its generation context jointly incorporates all extracted behavioral requirements. Stage~I then converts each requirement into a requirement-targeted scenario and stimulus and performs prefiltering to determine whether further constraint derivation is needed. If the stimulus for every behavioral requirement produces only a single cluster, no behavioral divergence is detected under the current candidate pool and generated stimuli. In this case, no surviving requirement family is produced, and the framework follows the fallback branch. One Verilog candidate from Stage~I is selected as the fallback candidate and passed directly to Stage~III. Otherwise, each requirement--scenario--stimulus combination that produces multiple behavioral clusters is retained as a surviving requirement--scenario--stimulus (RSS) family and forwarded to Stage~II for constraint derivation. After Stage~II, the Verilog candidates generated under the accumulated constraints, together with the last stimulus generated from the composite requirement, are passed to Stage~III. Stage~III then performs final candidate selection and output initialization to produce the final Verilog implementation.

\vspace{-4pt}
\subsection{Stage I: Structured Scenario--Stimulus Construction and Prefiltering}

This section introduces how Stage~I constructs structured stimuli and performs prefiltering, as illustrated in Figure~\ref{fig:stage1_prefilter}. Here, structured means that the generated scenarios and stimuli are written in an explicit, machine-usable form rather than as free-form text.

Starting from the original specification, LLMs first generate a fixed initial pool of \(N\) Verilog candidates, which is used for prefiltering and for extracting shared interface information. This interface information is necessary because the LLM must know the input and output port names in order to generate a correct simulation driver template. 

In parallel, SpecLens converts each behavioral requirement into a requirement-targeted scenario through the LLM-based Scenario Generator. A scenario supplements the requirement with the missing testing context implied by the specification. In many cases, the behavior described by a requirement begins in an intermediate state, while the stimulus must also specify how that state is reached. The scenario therefore provides a more complete description of the behavior to be tested. Each scenario is then processed by the LLM-based Stimulus Generator, which produces a structured stimulus template using the interface information. A Python-based Driver Generator then automatically converts this template into an executable Verilog driver file (\code{driver.v}) that applies inputs, handles clock and reset behavior, samples outputs, and logs traces.

 For the current requirement family, the generated \code{driver.v} is executed against the fixed initial Verilog candidate pool, and the resulting Verilog traces are grouped into the same cluster if they are identical. Candidates that fail to produce valid traces, typically because of
syntax or compilation errors, are automatically filtered out. The clustering result is then used to guide LLM-based stimulus refinement. For the same set of Verilog candidates, a stimulus that separates behaviors into more clusters provides more information about potential specification ambiguity. To quantify this effect, SpecLens computes behavioral entropy over the clusters induced by the stimulus. Let
\begin{equation}
\mathcal{G}(x)=\{G_1,G_2,\dots,G_K\}
\label{eq:clusters}
\end{equation}
denote the $K$ clusters induced by stimulus $x$, ordered in descending order by cluster size. Let $n_k$ denote the number of candidates in cluster $G_k$, and let
\begin{equation}
p_k=\frac{n_k}{M(x)}
\label{eq:pk}
\end{equation}
where $M(x)=\sum_{k=1}^{K} n_k$ denotes the number of Verilog candidates that produce valid traces under stimulus $x$. The behavioral entropy is then defined as
\begin{equation}
H(\mathcal{G})=-\sum_{k=1}^{K} p_k \log_2 p_k
\label{eq:entropy}
\end{equation}

Higher entropy indicates greater behavioral dispersion across clusters. Given the specification, a scenario, and its current stimulus, the LLM proposes a refined stimulus, and both the original and refined stimuli are evaluated on the same fixed initial Verilog candidates through the Python-based Clustering and Entropy Comparison (CEC). If entropy increases, refinement stops early, since the refined stimulus has already exposed greater behavioral dispersion. Otherwise, refinement continues for up to \(Q\) rounds.

Stage~I has two possible prefiltering outcomes. If all requirements produce only a single behavioral cluster, one Verilog candidate from the fixed initial pool is randomly selected and passed directly to Stage~III. Otherwise, any RSS combination that produces multiple clusters is kept as a surviving RSS family. These surviving families are then forwarded to Stage~II for constraint derivation.
\begin{figure*}[t]
    \centering
    \includegraphics[width=\textwidth]{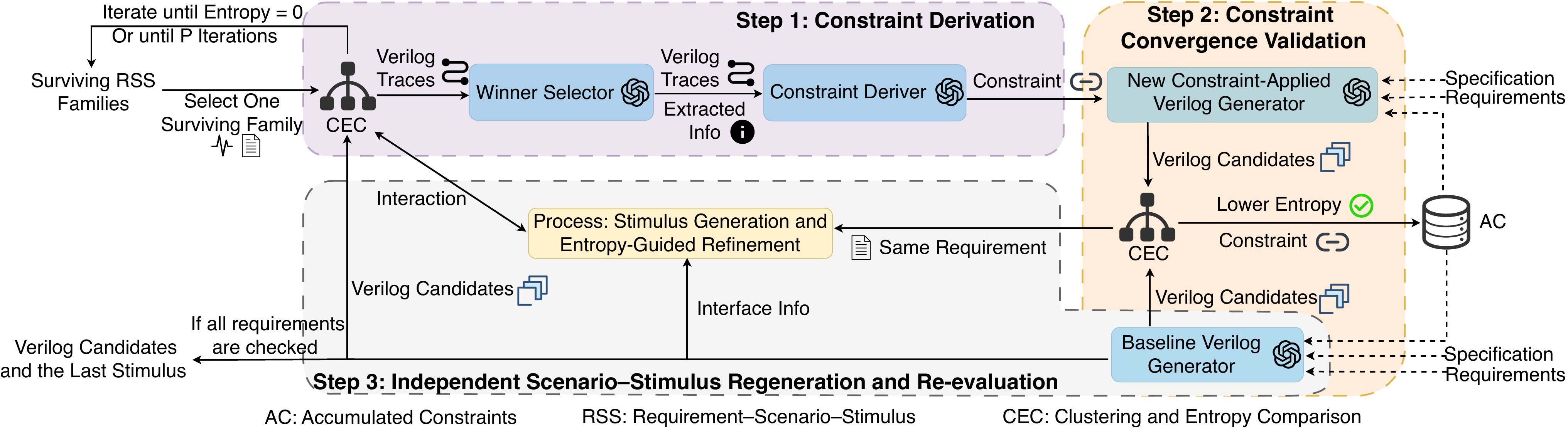}
    \Description{Stage~II: entropy-guided constraint discovery and iteration.}
    \caption{Stage~II: entropy-guided constraint derivation and iteration.}
    \label{fig:stage2_iteration}
\end{figure*}

\vspace{-4pt}
\subsection{Stage II: Entropy-Guided Constraint Derivation and Iteration}

Stage~II describes how constraints are iteratively derived from each RSS family and how entropy is used to validate them, as detailed in Figure~\ref{fig:stage2_iteration}. Each surviving RSS family from Stage~I enters an iterative constraint-derivation loop. This loop contains three steps.

\textbf{Step 1: Constraint Derivation}: This step derives disambiguation constraints from the input RSS family. At the beginning, the LLM-based Baseline Verilog Generator generates a batch of \(N\) Verilog candidates from the original specification, the behavioral requirement in the current RSS family, and the accumulated constraints for the same specification, if any. The CEC module then applies either the current stimulus from the RSS family or a newly refined stimulus from \textit{Step 3} to these candidates in order to cluster their resulting traces and compare different behaviors. If multiple clusters remain, the corresponding Verilog traces are forwarded for subsequent constraint derivation. If the candidates converge to a single cluster and the entropy becomes zero, the current RSS family is considered behaviorally converged under the current stimulus and the loop proceeds to the next RSS family.

If the entropy is nonzero, the top \(M\) clusters are selected, and their Verilog traces are used by the LLMs for winner selection and constraint derivation. Both steps are performed without access to the underlying Verilog code, so the framework focuses on differences in Verilog traces when judging whether a behavior satisfies the specification. This design avoids constraints tied to internal registers or implementation details. Instead, the resulting constraints supplement the specification with externally observable behavior.

The LLM-based Winner Selector uses only the original specification, the current requirement family, and the \(M\) Verilog traces to determine which trace is most consistent with the specification. The selected trace is then used as an anchor for subsequent constraint derivation, allowing the LLM-based Constraint Deriver to focus on the key behavioral difference among the \(M\) traces. Using the specification, the current requirement family, the anchor trace, and the \(M\) Verilog traces, it then derives a constraint to help later generated Verilog candidates avoid similar incorrect behavior.

For example, in the VerilogEval task \code{ece241\_2014\_q5a}, one common failure mode is a one-cycle-delayed response in which the output does not immediately reflect the first high serial input bit after reset. This error is identified by comparing Verilog traces under the same stimulus rather than inspecting internal code structure. Based on the observed mismatch, SpecLens derives an externally observable constraint.

\begin{example}[Derived constraint]
For \code{ece241\_2014\_q5a}, when \code{areset} is deasserted
and the module is processing serial input, on the positive clock edge where the first occurrence of \code{x=1} is applied after an
initial sequence of \code{0}s, the observed output \code{z} must be \code{1} in the same post-edge observation.
\end{example}

\textbf{Step 2: Constraint Convergence Validation}: A derived constraint from \textit{Step 1} is accepted only if it reduces behavioral entropy. To validate the constraint, the LLM-based New Constraint-Applied Verilog Generator takes it as additional semantic information and generates a new batch of Verilog candidates under the same specification and requirement context. These newly generated candidates are then evaluated by CEC under the same stimulus as before. If the entropy decreases, the constraint is accepted and added to the accumulated constraint set. Otherwise, it is rejected.

\textbf{Step 3: Independent Scenario–Stimulus Regeneration and Re-evaluation}: After \textit{Step 2}, the current requirement is sent back to the stimulus-generation and entropy-guided refinement process shown in Figure~\ref{fig:stage1_prefilter}, which independently regenerates a new scenario and refined stimulus. This regeneration improves the diversity and reliability of subsequent verification. If \textit{Step 2} accepts a new constraint, the LLM-based Baseline Verilog Generator in \textit{Step 3} produces a new batch of Verilog candidates under the updated accumulated constraint set, and these candidates are evaluated by CEC in \textit{Step 1} using the newly generated stimulus. If no new constraint is accepted, the Verilog candidates remain unchanged, and only the new stimulus is used in \textit{Step 1} to produce new Verilog traces.

The iterative loop for each RSS family continues until its entropy reaches zero or the maximum number of iterations \(P\) is reached. The framework then proceeds to the next surviving RSS family and repeats the same procedure. After all surviving RSS families have been processed, Stage~II outputs the current Verilog candidates together with the last stimulus generated from the composite requirement processed at the end of the requirement list.

\begin{figure}[t]
    \centering
    \includegraphics[width=\columnwidth]{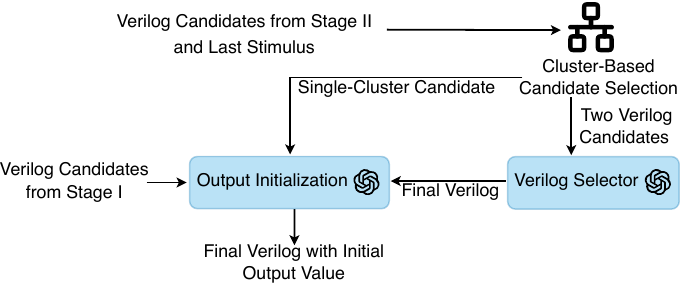}
    \Description{Stage III selects a final Verilog candidate from clustered candidates and initializes outputs according to the specification.}
    \caption{Stage~III: final Verilog selection and initialization.}
    \label{fig:stage3_finalization}
\end{figure}

\begin{table*}[t]
\centering
\caption{Comparison on VerilogEval v2.0 spec-to-RTL.}
\label{tab:verilogeval_v2_main}
\normalsize
\renewcommand{\arraystretch}{0.85}
\setlength{\tabcolsep}{14pt}
\begin{tabular}{llcccc}
\toprule
\multirow{2}{*}{Method} & \multirow{2}{*}{Model / Setting} & \multicolumn{2}{c}{Syntactic Correctness} & \multicolumn{2}{c}{Functional Correctness} \\
\cmidrule(lr){3-4} \cmidrule(lr){5-6}
 &  & pass@1 & pass@5 & pass@1 & pass@5 \\
\midrule
Self-planning~\cite{RTLLM_promptengineering} & o3-mini-medium & 94.9 & 99.7 & 78.0 & 88.2 \\
Two-shot prompting~\cite{vfocus} & o3-mini-medium & 98.5 & 99.9 & 82.6 & 90.0 \\
\textbf{SpecLens (ours)} & \textbf{o3-mini-medium} & \textbf{100.0} & \textbf{100.0} & \textbf{86.2} & \textbf{92.6} \\
DeepV~\cite{rag1} & GPT-5 Chat + RAG & 99.4 & 100.0 & 76.9 & 81.3 \\
VeriSeek~\cite{Reinforcement2} & Reinforcement Learning & 85.1 & 98.3 & 61.6 & 76.9 \\
SiliconMind-V1~\cite{finetune4} & Fine-tuning & -- & -- & 76.5 & 84.7 \\
VeriGRAG~\cite{rag3} & Fine-tuning + RAG & 99.4 & 100.0 & 62.2 & 69.5 \\
\bottomrule
\end{tabular}
\end{table*}

\vspace{-6pt}
\subsection{Stage III: Final Verilog Selection and Initialization}
This section presents how the final Verilog is produced in Stage~III, which has two input paths, as shown in Figure~\ref{fig:stage3_finalization}. If no requirement family survives Stage~I, one Verilog candidate from the fixed initial pool is passed directly to the initialization step, where the LLM assigns output initial values only according to the original specification. Otherwise, Stage~III receives the Verilog candidates generated in Stage~II together with the last stimulus. A Python-based cluster-selection module clusters and ranks the candidates according to their traces under this stimulus. If only one cluster remains, a candidate proceeds directly to initialization. If multiple clusters remain, one candidate is selected from each of the top two clusters, and the two candidates are sent to the LLM-based Verilog Selector, which selects the candidate that is more consistent with the specification. The selected Verilog is then initialized by an LLM according to the specification. This post-processing step is restricted to adding specification-supported output-initialization assignments and is intended to handle benchmark tasks with explicit default-output requirements.

\vspace{-6pt}
\section{Experimental Results}

\subsection{Setup}

Experiments were conducted on the original benchmark releases, including VerilogEval v2.0 spec-to-RTL
\cite{Revisiting_VerilogEval_promptengineering}, RTLLM v1.1 \cite{RTLLM_promptengineering}, and RTLLM v2.0 \cite{RTLLM2}. For syntax checking and functional verification, we used Icarus Verilog~\cite{iverilog} and Verilator~\cite{verilator}. Functional correctness was measured by running the final Verilog generated in Stage~III against the benchmark testbench.

The maximum number of stimulus refinement rounds was set to $Q=3$ in both Stage~I and Stage~II. We set the maximum number of
iterations per requirement in Stage~II to $P=3$. The number of fixed initial Verilog candidates generated in Stage~I was set to $N=8$, and the number of Verilog candidates generated in each Stage~II round for entropy evaluation was also set to $N=8$. The primary evaluation metric is pass@k.
Let $n$ denote the number of sampled candidates for a problem and $c$ denote the number of functionally correct candidates among them. The pass@k score is computed as
\begin{equation}
\mathrm{pass@}k = \mathbb{E}_{\mathrm{problems}}
\left[
1 - \frac{\binom{n-c}{k}}{\binom{n}{k}}
\right]
\label{eq:passk}
\end{equation}
This metric estimates the probability that at least one correct solution appears among $k$ sampled candidates. We set $n=20$.

In Step~1 of Stage~II, we set CEC to select Verilog traces only from the top \(M=2\) clusters. In a post hoc analysis of SpecLens failure cases, we examine the risk of restricting comparison to the top two clusters. Under our main setting with o3-mini-medium, the probability that a round produces more than two clusters and that neither of the top two clusters contains a functionally correct Verilog implementation is only 2.73\%. For o3-mini-high, the corresponding probability averages 15.8\%. This suggests that stronger models more often produce correct implementations as low-frequency outliers. Developing more robust cluster-selection mechanisms for such settings is left for future work.

Table~\ref{tab:verilogeval_v2_main} compares SpecLens with representative prompting, reinforcement learning, fine-tuning, and retrieval-based methods on VerilogEval v2.0 spec-to-RTL. The two-shot in-context learning baseline uses two examples from \cite{vfocus}: a combinational XOR gate and a sequential 8-bit registered incrementer. The self-planning prompt is taken from \cite{RTLLM_promptengineering}. We also ran these two baselines with o3-mini-medium. The comparison also includes retrieval-based methods, such as DeepV \cite{rag1}. For comparison with fine-tuning-based methods, we include SiliconMind-V1~\cite{finetune4} and VeriGRAG~\cite{rag3}. SiliconMind-V1 is a recent fine-tuned Verilog model, while VeriGRAG combines fine-tuning with graph-embedding-based retrieval. It is worth noting that DeepV and VeriSeek report results on the original VerilogEval benchmark, whereas our main evaluation uses VerilogEval v2.0. We therefore include these results only as cross-version references, rather than as directly comparable baselines.

\vspace{-7pt}
\subsection{Results and Comparison}

 As shown in Table~\ref{tab:verilogeval_v2_main}, SpecLens achieves the best overall performance, reaching 100.0\% for both syntactic pass@1 and pass@5, as well as 86.2\% and 92.6\% for functional pass@1 and pass@5, respectively. It outperforms both self-planning and two-shot prompting. A similar advantage is observed over retrieval-augmented methods such as DeepV. These results suggest that directly identifying specification-specific uncertainty is more effective than simply adding externally provided information.

Furthermore, we compare against VRank~\cite{vrank} and VFocus~\cite{vfocus}. Because the original VRank paper does not report results for o3-mini, we use the o3-mini-high VRank result reported in the follow-up VFocus study~\cite{vfocus}. For a fair comparison, SpecLens, VRank, and VFocus are all compared under the same o3-mini-high setting. Under this setting, VRank achieves a functional pass@1 of 77.4\%, and VFocus improves it to 85.6\%. SpecLens further reaches 89.4\%. These results suggest that behavioral divergence is useful not only as a back-end mechanism for candidate selection, but also as a way to identify where the LLM is uncertain about the specification and to provide targeted constraints for those ambiguous parts.

\begin{figure}[t]
    \centering
    \includegraphics[width=\columnwidth]{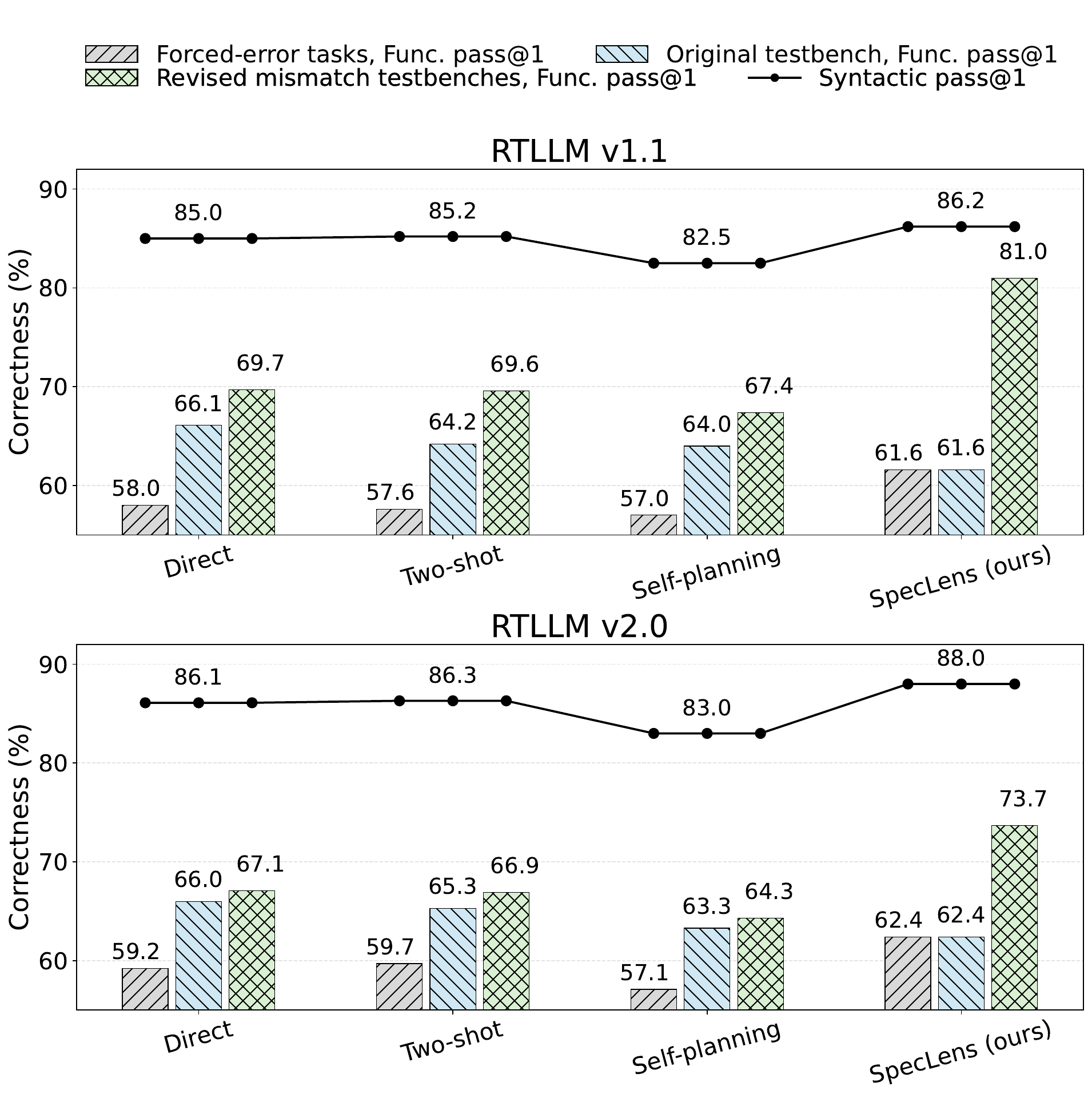}
    \Description{Comparison of syntactic and functional pass@1 on RTLLM v1.1 and RTLLM v2.0 under original, forced-error, and revised-testbench settings.}
    \caption{Comparison of syntactic and functional pass@1 on RTLLM v1.1 and RTLLM v2.0 under three settings.}
    \label{fig:rtllm_main}
\end{figure}

Figure~\ref{fig:rtllm_main} reports the main results on RTLLM v1.1 and RTLLM v2.0. The line plots show syntactic pass@1, and the bar plots show functional pass@1. Three baselines are included: direct prompting, two-shot prompting~\cite{vfocus}, and self-planning~\cite{RTLLM_promptengineering}, all under the same base model setting, o3-mini-medium. 

Under the originally released testbenches, SpecLens achieves higher syntactic pass@1 than the baselines, but functional pass@1 is lower on both datasets, as shown by the green bars. A manual audit reveals specification--testbench mismatches in many tasks for which the baselines are counted as correct but SpecLens is counted as incorrect. A representative example is \texttt{alu}. In the released specification, the final sentence states that the LUI behavior should use the upper 16 bits of input \texttt{a}. However, the baselines still often generate Verilog that matches the released testbench rather than this local specification statement. To further verify this effect, we intentionally replaced the local statement with the following version: ``For the LUI operation, the upper 8 bits of input \texttt{a} are concatenated with 8 zeros to form the result.'' Even after this explicit local change, the three baselines still frequently produced the original benchmark-aligned implementation. This suggests that the baselines are strongly influenced by pretraining and tend to preserve previously learned RTL patterns, such as standard LUI rules, instead of following local specification changes. As a result, an implementation may pass the released testbench without necessarily conforming to the current specification.

Specification--testbench mismatches are identified in six tasks in RTLLM v1.1: asyn\_fifo, multi\_pipe\_8bit, parallel2serial, alu, traffic\_light, and width\_8to16. RTLLM v2.0 contains the same six cases plus one additional case, fixed\_point\_substractor. Based on these identified mismatch cases, we further report two additional evaluation settings by changing only how these tasks are counted, while leaving all other tasks unchanged. In both settings, the evaluated Verilog candidates are exactly the same as those already produced and tested under the originally released testbenches, and no new Verilog is generated. Therefore, syntactic pass@1 is unchanged. SpecLens has the highest syntactic pass@1, at 86.2\% on RTLLM v1.1 and 88.0\% on RTLLM v2.0. In the forced-error mismatch tasks setting, all seven tasks are counted as functionally incorrect for all four methods; this result is shown by the orange bars. Notably, SpecLens has the same functional pass@1 under the forced-error setting and under the original testbench setting, 
because none of these seven tasks passes the originally released testbench under SpecLens. In the revised-testbench setting, the original specification and testbench for each of these seven tasks are jointly provided to the LLM to generate a revised testbench, which is then manually reviewed to ensure better consistency with the specification; this result is shown by the purple bars. The goal of this revision is to remove clear mismatches between the specification and the released testbench. For example, in \texttt{asyn\_fifo}, where the specification gives conflicting definitions of \texttt{DEPTH}, the revised testbench checks only the external FIFO behaviors shared by both interpretations, thereby avoiding dependence on either ambiguous interpretation. Under the revised testbenches, SpecLens reaches 81.0\% functional pass@1 on RTLLM v1.1 and 73.7\% on RTLLM v2.0.

These results suggest that direct prompting more readily preserves standard RTL templates and architectural priors acquired during pretraining. Few-shot prompting and self-planning can enrich the input context, but they do not force the model to prioritize a modified local statement over stronger benchmark-aligned priors. By contrast, SpecLens is designed to surface and reinforce such local behavioral statements, making its outputs more likely to depart from pretrained answers when the specification changes.

The derived disambiguation constraints are transferable across models. On VerilogEval v2.0, using them as additional guidance improves GPT-4o-mini's functional pass@1 ratio by 6.84\% relative to using the original specification alone, reaching \(\mathrm{pass@}1 = 54.7\%\).

\vspace{-7pt}
\subsection{Ablation Study}

\emph{Effect of Prefiltering}. Stage I prefiltering prevents unnecessary requirements from entering downstream constraint derivation. Under our main setting on VerilogEval v2.0, it reduces token consumption by 58.7\% to 28.7M tokens and runtime by 72.7\% to 20,527 seconds, with nearly unchanged functional correctness.

\emph{Effect of Entropy-guided Constraint Validation}. On VerilogEval v2.0, removing entropy filtering reduces functional pass@1 from 86.2\% to 84.0\%. Without it, newly generated constraints are applied directly even when they are redundant or partially misleading.

\emph{Importance of Explicit Requirement Extraction}. When the requirement extractor and aggregator are removed and scenarios are generated directly from the specification, functional pass@1 on VerilogEval v2.0 decreases from 86.2\% to 82.9\%. Scenario coverage also decreases, from 891 generated scenarios with requirement extraction to 781 without it.

\emph{Effect of Stimulus Refinement}. On VerilogEval v2.0, the average stimulus length in Stage~I increases from 3.85 to 4.32 steps. In Stage~II, the average stimulus length increases from 7.34 to 8.71 steps, a gain of about 19\%. The task-level probability that stimulus refinement produces an entropy-improving stimulus is 3.84\%.

\vspace{-7pt}
\section{Conclusion}

This paper presents SpecLens, an automated framework for LLM-based Verilog generation that derives disambiguation constraints from behavioral divergence among candidate implementations. Without relying on large training corpora or external Verilog libraries, SpecLens is more general and scalable than fine-tuning and retrieval-based methods. Experiments show that it improves functional correctness on VerilogEval v2.0 and produces more specification-faithful Verilog on RTLLM under flawed or locally modified specifications.

\end{document}